\documentclass[a4paper]{article}

\usepackage{INTERSPEECH2019}
\usepackage{multirow}
\usepackage{graphicx}
\usepackage{xcolor}
\usepackage{subfigure}
\usepackage{verbatim}
\usepackage{color,soul}

\title{Overcoming label noise in audio event detection using sequential labeling}
\name{Jae-Bin Kim$^1$, Seongkyu Mun$^2$, Myungwoo Oh$^2$, Soyeon Choe$^2$, Yong-Hyeok Lee$^1$,  Hyung-Min Park$^1$}
\address{$^{1}$Department of Electronic Engineering, Sogang University, South Korea\\ $^{2}$Naver Corporation, South Korea}

\email{hpark@sogang.ac.kr}

\begin{document}

\maketitle
\begin{abstract}
This paper addresses the noisy label issue in audio event detection (AED) by refining strong labels as sequential labels with inaccurate timestamps removed.
In AED, strong labels contain the occurrence of a specific event and its timestamps corresponding to the start and end of the event in an audio clip. The timestamps depend on subjectivity of each annotator, and their label noise is inevitable. Contrary to the strong labels, weak labels indicate only the occurrence of a specific event. They do not have the label noise caused by the timestamps, but the time information is excluded. To fully exploit information from available strong and weak labels, we propose an AED scheme to train with sequential labels in addition to the given strong and weak labels after converting the strong labels into the sequential labels. Using sequential labels consistently improved the performance particularly with the segment-based F-score by focusing on occurrences of events. In the mean-teacher-based approach for semi-supervised learning, including an early step with sequential prediction in addition to supervised learning with sequential labels mitigated label noise and inaccurate prediction of the teacher model and improved the segment-based F-score significantly while maintaining the event-based F-score.


\end{abstract}

\section{Introduction}
\label{intro}


Audio event detection (AED) refers to the task of recognizing when and which audio events occur in an audio recording~\cite{turpault2019sound}. In order to train an AED system, one may need a dataset with strong labels that annotate timestamps corresponding to the start and end of event occurrences in addition to their presence or absence. Unfortunately, annotating the strong labels is too laborious to develop a large-sized strongly labeled dataset. Furthermore, many kinds of audio events \textit{e.g.} footsteps, wind blowing, or burning fire, etc, have a vague start or end in time, and some labeled datasets have no verification between inter-annotators, as mentioned in the previous DCASE challenge \cite{mesaros2016tut}. Therefore, the timestamps depend on the subjectivity of each annotator, which are frequently inaccurate or noisy. Therefore, 
a number of recent audio event detection (AED) researches have used weakly labeled data, such as AudioSet \cite{gemmeke2017audio} and FSD \cite{fonseca2017freesound}, where do not have the timestamps. 
In particular, training on weakly labeled data have been widely researched using multiple instance learning recently \cite{turpault2019sound, wang2019comparison, serizel2018large}.

\begin{figure}
    \centering
    \includegraphics[width=0.89\columnwidth]{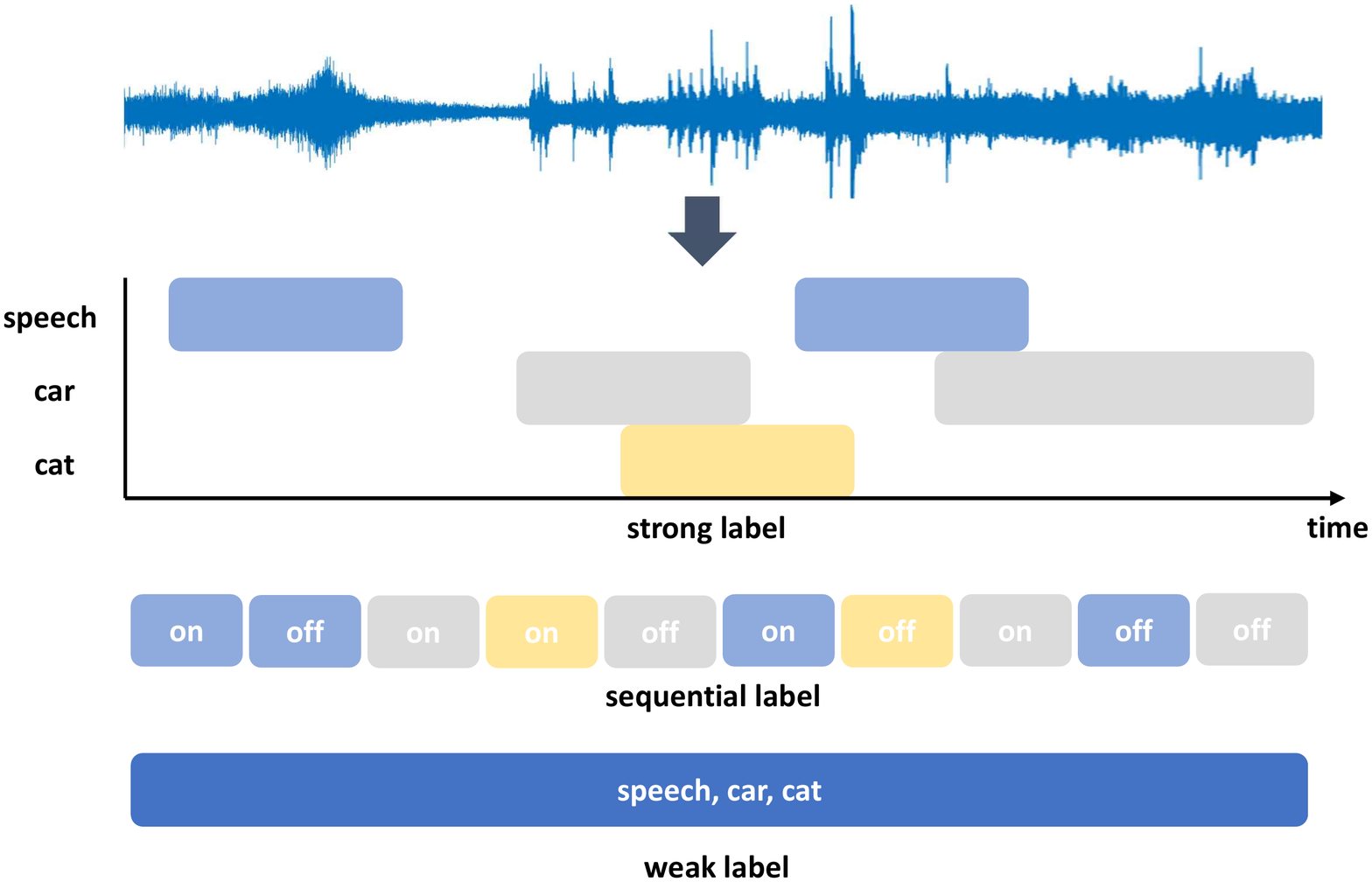}
    \caption{Comparison of strong, sequential, and weak labeling.}
    \label{fig:labeltype}
\end{figure}

Despite various training approaches, the weak label has an apparent limitation that there is no time information. To improve the performance of AED systems based on the weak labels only, additional use of sequential labels describing temporal sequential relationship of events was proposed by \cite{wang2019connectionist}, where the start and end points of audio events were mapped as individual label symbols. As shown in Figure \ref{fig:labeltype}, the sequential label provides only the sequence of event boundaries instead of indicating their timestamps. The author focused on the possibility of connectionist-temporal-classification(CTC)-based framework being applied to AED. Given strong, weak, and sequential labels, the author trained three models individually using each label set and training with sequential labels showed the mid-level performance between strong and weak labels. 

Although strong labels were available for all data in~\cite{piczak2015esc,temko2006clear,nakamura1999data}, 
very limited data are annotated 
by strong labels in many practical datasets such as the DCASE challenges~\cite{turpault2019sound}. In this situation, strong labels are very important to match a specific sound to its event class in an AED model from the scratch even though their timestamps are noisy. In addition, it is noteworthy that sequential labels are noise-robust information which can be obtained from strong labels. To fully exploit information from available strong and weak labels, we propose an AED scheme to train with sequential labels in addition to the given strong and weak labels after converting the strong labels into the sequential labels.
Since noise-robust information of the strong labels are refined in the sequential labels, the sequential labels may provide consistent cues to train an AED model. In addition, unlike weak labels, the sequential labels contain temporal sequential relationship of events that is useful to guide the model, which may result in performance improvement when using the sequential, strong, and weak labels simultaneously.


Furthermore, the proposed approach can be applied for semi-supervised learning, such as teacher-student learning.  Like the annotator's error, noise (or inaccuracy) also exists in pseudo-labels generated from the teacher model during training phase
, and the label noise is more severe on strong labels than weak labels. Therefore, we also propose an approach to use teacher-student learning based on sequential labels which are more noise-robust than the strong labels and more informative than the weak labels.

\section{Related Works}
The main topic of this paper is how to use sequential label for training better model.
To train the sequential label without temporal alignment, preliminary researches have been conducted on CTC in AED \cite{wang2019connectionist, wang2017first}. Based on its effective application, we also used CTC and Connectionist Temporal Localization (CTL) frameworks \cite{wang2019connectionist} for the sequential label training. 



\subsection{Connectionist temporal classification}
CTC was first proposed in speech recognition \cite{graves2006connectionist}. Rather than frame-wise supervision, the CTC framework only requires phone sequences of the training utterances without an alignment between phonemes and frames.
CTC considers all possible phone sequence cases through a phone combination with blank labels. To merge them into a single output, many-to-one mapping is used for removing blank or repeated labels.
Using many-to-one mapping, the total output probability can be derived as the sum of the all possible probabilities.

As for the output label of AED, the author of \cite{wang2017first} suggested to use boundaries of sound events. For example, if the audio clip contains speech, car and cat sound as shown in Figure 1, sequence-label is ``speech-start, speech-end, car-start, cat-start, car-end, speech-start, cat-end, car-start, speech-end, car-end" as ground truth. Therefore, $n$-class system has $2n + 1$ output nodes, two for each sound event and one for blank token.
Through various experiments, the author showed the feasibility of novel labeling structure for AED.


\subsection{Connectionist temporal localization}

After the preliminary AED research using CTC, the author of \cite{wang2019connectionist} proposed the CTL system to address a issue called ``peak clustering". As shown in Figure 2 in \cite{wang2019connectionist}, the CTC system tends to predict on/offset repeatedly next to each other for a long single event input. This consecutive event boundaries made misdetected event clusters, and was named ``peak clustering". The author analyzed that this issue is mainly due to the multiple purpose of using a blank label. The blank label in CTC serves two purpose: (1) emitting ``no event" at a frame and (2) separation of same event repetition. These inconsistent objectives interfere with training blank label properly, therefore occurs frequent blank-output which can lead to consecutive event peaks. To address this issue, the author proposed the CTL framework that eliminates both blank and on/offset label in the output layer of model.

In the CTL framework, the model estimates the probability of event classes for each frame, and derives the event boundary probability by estimating on/offset using rectified delta operator. Let $y_t(E)$ is the probability that event $E$ being active at frame $t$, and $z_t(\acute{E})$ and $z_t(\grave{E})$ are the probability of on/offset of the event \emph{E} at frame \emph{t}. $z_t(\acute{E})$ and $z_t(\grave{E})$ are calculated as follows :
\begin{equation} 
    z_t(\acute{E}) = \max[0, y_t(E) - y_{t-1}(E)].
\end{equation}
\begin{equation} 
    z_t(\grave{E}) = \max[0, y_{t-1}(E) - y_t(E)].
\end{equation}





CTL assumes the probabilities of different event boundaries at the same frame as mutually independent instead of mutually exclusive. 
In this way, the probability of no event boundaries  occurring at frame is calculated by:

\begin{equation}\label{eq:epsilon}
    \epsilon_t = \textstyle\prod_{l}[1-z_t(l)],
\end{equation}
where $l$ goes over all event boundaries. 
The probability of emitting a single event boundary $l$ at frame $t$ is then:

\begin{equation} \label{eq:p_sl}
    p_t(l) =\textstyle z_t(l) \cdot \prod_{l'\neq l}[1-z_t(l')].
\end{equation}
If we define

\begin{equation} \label{eq:delta}
    \delta_t(l) =\frac{z_t(l)}{1-z_t(l)},
\end{equation}
then, we can get 
\begin{equation}\label{eq:p_rd}
    p_t(l) = \textstyle\epsilon_t \cdot \delta_t(l).
\end{equation}
From Eq. (\ref{eq:epsilon}), the blank label can be eliminated, and based on the modification of many-to-one mapping function and forward algorithm of CTC, the blank label is no longer used for separating repetition. More details on CTL can be found in \cite{wang2019connectionist}.

The CTL can allow the multiple event occurrence at the same frame. However, this is rare in practice and the result of \cite{wang2019connectionist} showed AED performance deterioration. Therefore, we assume that there is no event co-occurrence in this work.



\section{Proposed Approach}
We propose to use sequential labeling for two-types of label noise : strong label noise by human annotator and inaccuracy of strong prediction by a teacher model.

\subsection{Sequential labeling using noisy label}
Sequential labeling is quite intuitive. As shown in Figure 1, on/offset of each event segment are extracted from strong label, and sorted in chronological order. We used the CTC/CTL framework for training the  sequential label.
The AED system can be trained only with sequential label, but like recent challenges \cite{turpault2019sound,serizel2018large,fonseca2019audio}, other types of label also can be used jointly for training. If we use the CTL framework for sequential label, computing frame-wise probabilities is same way as other labels, therefore the strong, weak, and sequential labeling system can be combined with no additional effort. The weight for individual losses was heuristically selected through intensive experiments, \textit{e.g.} strong : weak : sequential label = 4 : 2 : 1. We verified the performance comparison in detail when the sequential or strong label is used alone and with other types of labels.

\subsection{Expansion to semi-supervised learning}

To exploit the unlabeled data effectively, sequential labeling is applied to the teacher-student learning. In this work, we used a mean-teacher-based approach \cite{tarvainen2017mean} widely used in AED \cite{turpault2019sound, jiakai2018mean}. The main concept of this approach is averaging model weights over training steps to produce a better model. The teacher model updated by a moving average of student model weights, and the consistency cost, in addition to the classification cost, is used for comparing the prediction between the student and teacher models. After the training, the teacher model is used for evaluation. 

The consistency cost for both strong and weak-prediction has already been used for AED \cite{jiakai2018mean}. This strong-prediction on unlabeled data by teacher model is similar to labeling process by human annotator.
However, using three-types of label (strong, weak and sequential) from beginning of training showed unstable learning curve, since prediction inaccuracy in model training is much worse than label noise caused by annotator's subjectivity. Therefore, we developed mean-teacher learning scheme for using sequential label. The algorithm is described by pseudo-code in Figure 2. We used Mean Squared Error as strong and weak consistency cost function and CTL as sequential consistency cost function. Also, we used sigmoid ramp-up function for weighting consistency cost as proposed in \cite{tarvainen2017mean}. The maximum value of ramp-up function was set to 1 through intensive experiments. 

From the beginning to half-point of training schedule, sequential label is used for loss instead of strong label, and vice versa for rest of the schedule. We found that training performance also showed unstable when strong label used first. We consider this result is due to minimizing sequential-consistency loss is relatively easier than strong-consistency. In other word, the strong-prediction difference between student and teacher model is huge and inaccurate in the beginning of training. Therefore, using sequential-prediction can mitigate instability of model training in early stage.
\begin{figure}
    \centering
    \includegraphics[width=0.92\columnwidth]{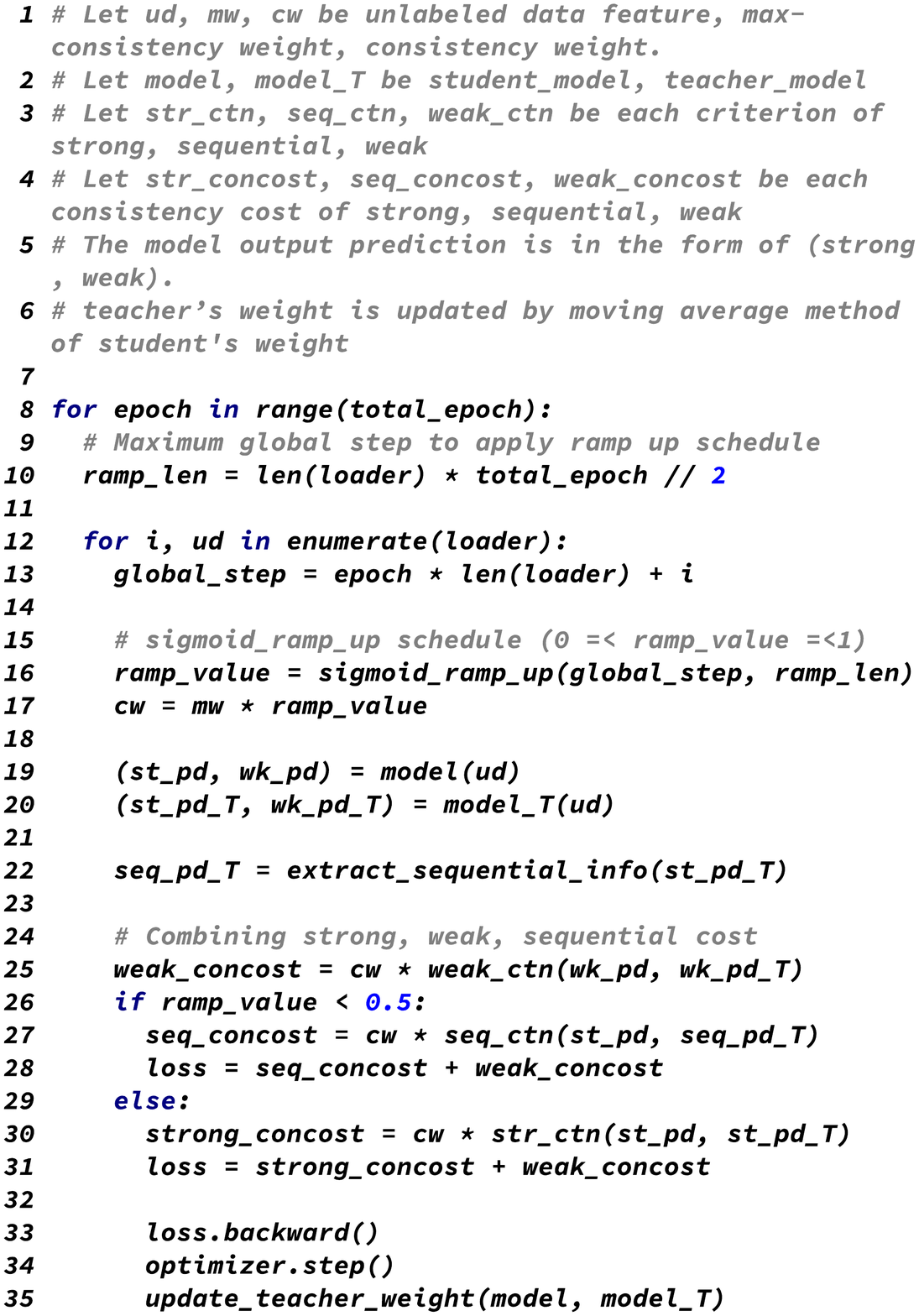}
    \caption{Python-style pseudo-code for the semi-supervised learning with sequential label.}
    \label{fig:pseudocode}
\end{figure}

\section{Experiments and Results}
\begin{figure*}
    \centering
    \subfigure[Strong label supplementation using sequential label.]{
    \includegraphics[width=1\columnwidth, height=1.8in]{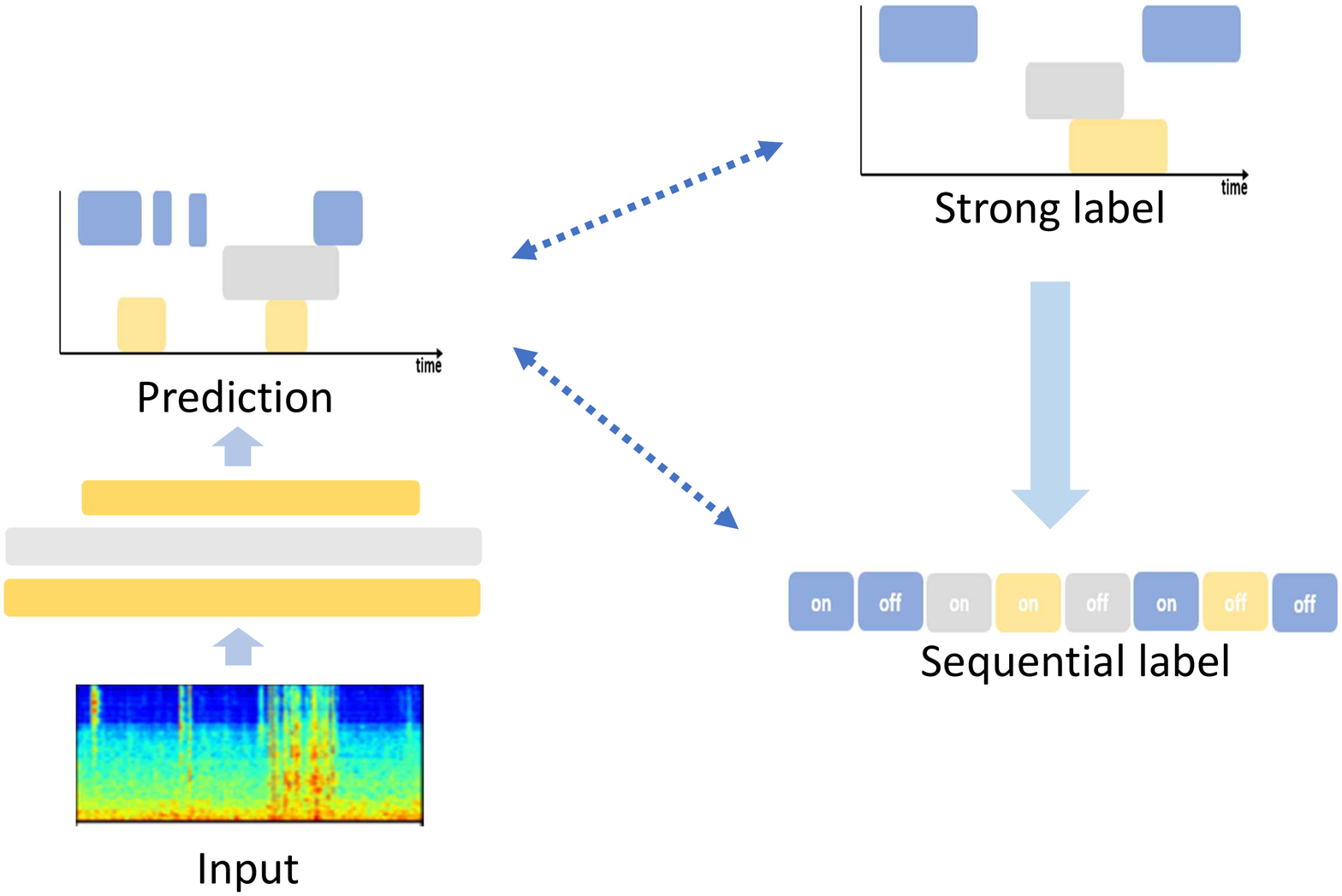}
    \label{fig:Method1}
    }
    \subfigure[Sequential mean-teacher.]{
    \includegraphics[width=1\columnwidth, height=1.8in]{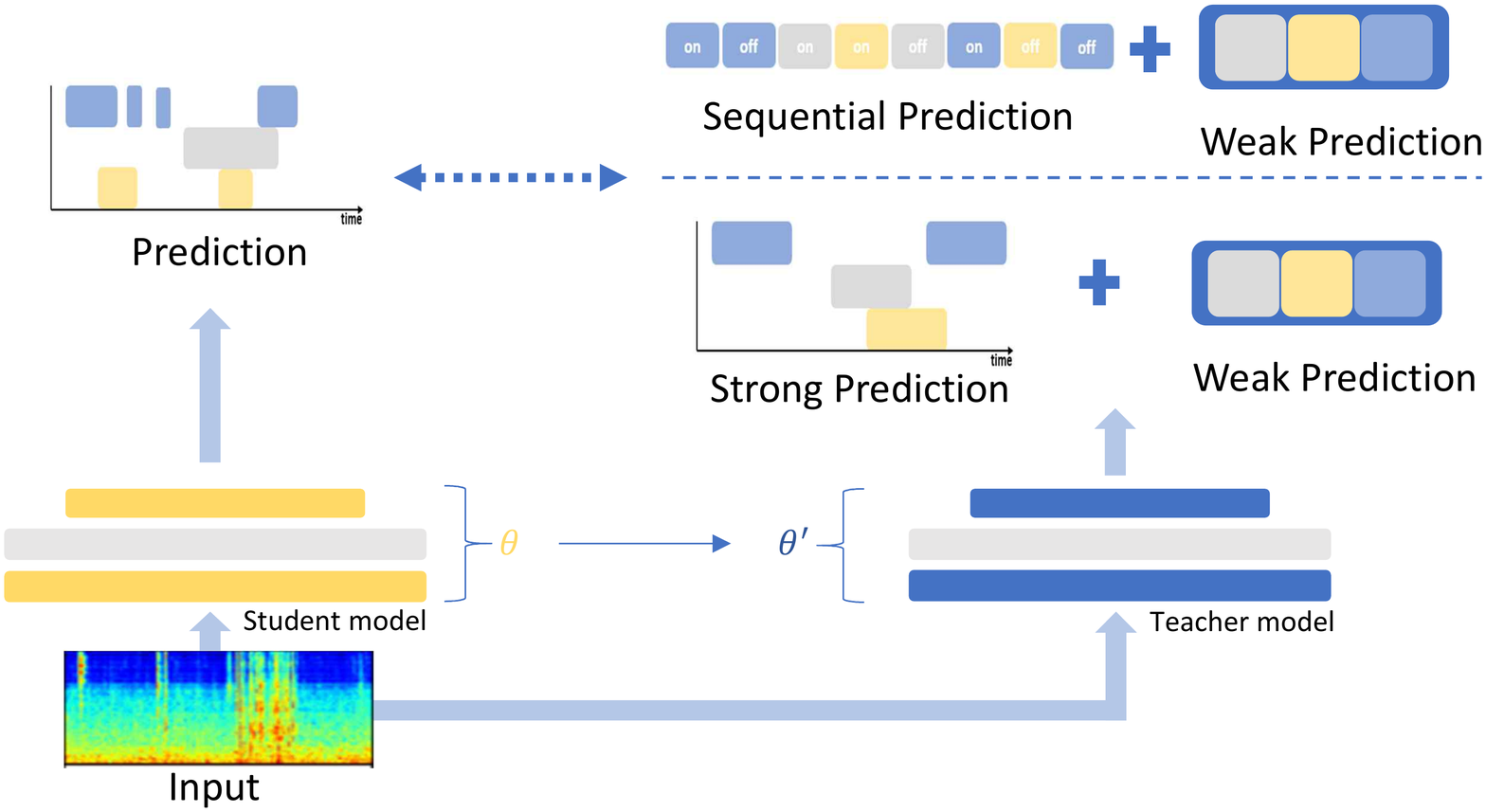}
    \label{fig:Method2}
    }
    \caption{Our proposed method  description.}
    \label{fig:Exschematic}

\end{figure*}

In this section, we describe our experiment setup and results. Our proposed method was described in Figure \ref{fig:Exschematic}, evaluated and compared with the DCASE 2019 baseline. 

\subsection{Experimental setup}
We have used the datasets of DCASE 2019 Task 4 and DCASE 2016 Task 3. 
The DCASE 2019 Task 4 dataset to classify 10 sound classes in the domestic environment consists of 2,045 strongly-labeled synthetic data, 1,578 weakly-labeled data, and 14,412 unlabeled data. The weakly-labeled and unlabeled data are real-recorded data.
The DCASE 2016 Task 3 dataset to classify 17 sound event classes consists of real-recorded data with strong annotations. The strong annotations were conducted by two research assistants trained through several example recordings, and more information is described in \cite{mesaros2016tut}.

We used the DCASE 2019 Task 4 Training set and the DCASE 2016 Task 3 Development set for training, and the DCASE 2019 Task 4 Public Evaluation set and the DCASE 2016 Task 3 Evaluation set for evaluation. 
To evaluate our proposed method, the DCASE 2019 Task 4 Baseline \cite{turpault2019sound} was used as our baseline model, and the evaluation metrics were event-based F-score (macro average) and segment-based F-score (macro average) computed using the sed\_eval library \cite{mesaros2016metrics}.

\subsection{Experimental results}

\subsubsection{CTC vs CTL}
\label{sec:CTCCTL}
We conducted an experiment to compare the performance of models trained by CTC and CTL when using strong or sequential labels. In case of using the CTC, we trained our model by multi-conditional learning because the model should predict the event boundary probabilities. Table \ref{tab:CTCvsCTL} summarizes the F-scores of models trained by the CTC and CTL. For comparison, a model trained by strong labels only is also evaluated. The performance of using the CTC only was significantly lower than the others, which might be due to the ``peak clustering" problem. Using the CTL showed comparable performance with the loss based on the strong labels only. When using both the sequential and strong labels, training based on the CTC provided a higher segment-based F-score than the case using either one while CTL-based training improved both the event- and segment-based F-scores. It demonstrates the effectiveness of the proposed method to supplement the strong labels with the sequential labels without causing 
the ``peak clustering" problem even though the sequential labels were obtained from the strong labels. Therefore, we conducted subsequent experiments by using the CTL as the sequential loss instead of the CTC.



\begin{table}[th!]
\centering

\caption{F-scores (\%) of models trained with losses based on CTC/CTL or strong labels  for the DCASE 2019 Task 4 Public Evaluation set.} 
\label{tab:CTCvsCTL}
\begin{tabular}{|c|c|c|}
\hline
\begin{tabular}[c]{@{}c@{}}Training Loss \\ \end{tabular} & \begin{tabular}[c]{@{}c@{}}Event-based \\ F-score \end{tabular} & \begin{tabular}[c]{@{}c@{}}Seg.-based \\ F-score \end{tabular} \\ \hline
CTC only                                                       & ~1.34                                                                   & ~9.15                                                                     \\ \hline
CTL only                                                       & 17.27                                                                   & 38.24      \\ \hline
strong-label loss only                                                     & 15.24                                                                   & 41.62                                                                     \\ \hline
strong-label and CTC losses                                                       & 14.21                                                                   & 42.55                                                                     \\ \hline
strong-label and CTL losses                                                       & \textbf{17.71}                                                                   & \textbf{44.19}                                                                     \\ \hline
\end{tabular}
\end{table}

\subsubsection{Sequential labeling for noisy strong labels}
\label{sec:method1}
We evaluated our proposed approach based on sequential labels refined from strong labels, using the DCASE 2016 Task 3 Development set or strongly labeled data in the DCASE 2019 Task 4  Training set for training.

Table \ref{tab:2016result} shows the F-scores on the DCASE 2016 Task 3 dataset. 
Considering that the model trained with both strong and sequential labels achieved better performance than that trained with the strong labels only, sequential information was helpful for improving the performance. In particular, performance improvement in the segment-based F-score was greater than in the event-based F-score since the sequential information focused on occurrences of events to provide consistent and noise-robust cues useful for training. 

\begin{table}[th!]
\centering
\renewcommand{\arraystretch}{0.9}
\caption{F-scores (\%) of models trained with strong labels with and without sequential labels on the DCASE 2016 Task 3 dataset.}
\label{tab:2016result}
\begin{tabular}{|c|c|c|}
\hline
\begin{tabular}[c]{@{}c@{}}Labels for training \end{tabular} & \begin{tabular}[c]{@{}c@{}}Event-based \\ F-score \end{tabular} & \begin{tabular}[c]{@{}c@{}}Seg.-based \\ F-score \end{tabular} \\ \hline
strong labels only                                                    & 6.17                                                                    & 17.87                                                                     \\ \hline
strong and sequential labels                                             & \textbf{7.86}                                                           & \textbf{23.38}                                                            \\ \hline
\end{tabular}
\end{table}

\begin{table}[h!]
\renewcommand{\arraystretch}{0.9}
\caption{F-scores (\%) of models trained with various combination of strong, weak, and sequential labels when using strongly labeled data in the DCASE 2019 Task 4 Training set for training. Given strong labels were converted into weak and sequential labels. Note that weakly labeled data from the DCASE training set are not used in this experiment.}
\centering
\label{tab:M1_result_woMT}
\resizebox{\columnwidth}{!}{%
\begin{tabular}{|c|c|c|c|c|}
\hline
\multicolumn{3}{|c|}{Label for training} &
  \multicolumn{1}{c|}{\multirow{2}{*}{\begin{tabular}[c]{@{}c@{}}Event-based\\ F-score\end{tabular}}} &
  \multicolumn{1}{c|}{\multirow{2}{*}{\begin{tabular}[c]{@{}c@{}}Seg.-based\\ F-score\end{tabular}}} \\ \cline{1-3}
\multicolumn{1}{|c|}{strong} &
  \multicolumn{1}{c|}{weak} &
  \multicolumn{1}{c|}{sequential} &
  \multicolumn{1}{c|}{} &
  \multicolumn{1}{c|}{} \\ \hline
\checkmark          &           &                 & 15.24             & 41.62             \\ \hline
            & \checkmark        &                 & ~7.63             & 33.80             \\ \hline
            &           & \checkmark              & 17.27            & 38.24             \\ \hline
\checkmark          & \checkmark        &                 & 12.54             & 38.86             \\ \hline
\checkmark          &           & \checkmark              & \textbf{17.71}            & 44.19             \\ \hline
            & \checkmark        & \checkmark              & 13.72             & 43.14             \\ \hline
\checkmark          & \checkmark        & \checkmark              & 17.54            & \textbf{48.62}             \\ \hline
\end{tabular}%
}

\end{table}

Table \ref{tab:M1_result_woMT} summarizes the F-scores of models trained with various combination of strong, weak, and sequential labels when using strongly labeled data in the DCASE 2019 Task 4 Training set for training. To conduct experiments with weak or sequential labels, given strong labels were converted into weak and sequential labels. Training with the weak labels only containing the least information provided the worst performance. Consistent with the results in Table~\ref{tab:CTCvsCTL}, training with the sequential labels had a lower segment-based F-score and a higher event-based F-score than that with the strong labels because of label noise in the strong labels. 
Adding the sequential labels to the strong or weak labels consistently improved both the event- and segment-based F-scores. As mentioned in Section~\ref{intro}, that is because the sequential labels provided consistent cues to train the model by refining noise-robust information of the strong labels and retained temporal sequential relationship of events that the weak labels did not have. 
In contrast, the model trained with the strong and weak labels showed the mid-level performance between training with the strong and weak labels only because 
the weak and strong labels were so different in label characteristics that they could not supplement each other and had independently influenced the model.
Moreover, the model trained with all the strong, weak, and sequential labels achieved the best segment-based F-score and an event-based F-score comparable with the model trained with both the strong and sequential labels by fully exploiting available information.

\begin{table}
\renewcommand{\arraystretch}{0.9}
\caption{F-scores (\%) on the mean-teacher-based approach for semi-supervised learning when using the DCASE 2019 Task 4 Training set as training data.}

\label{tab:Method2}
\resizebox{\columnwidth}{!}{%
\begin{tabular}{|c|l|l|c|c|}
\hline
\multicolumn{3}{|c|}{\multirow{2}{*}{Training method}} &
  \multicolumn{1}{c|}{\multirow{2}{*}{\begin{tabular}[c]{@{}c@{}}Event-based\\ F-score\end{tabular}}} &
  \multicolumn{1}{c|}{\multirow{2}{*}{\begin{tabular}[c]{@{}c@{}}Seg.-based\\ F-score\end{tabular}}} \\
\multicolumn{3}{|c|}{} & \multicolumn{1}{c|}{} & \multicolumn{1}{c|}{} \\ \hline
\multicolumn{3}{|c|}{conventional approach}                                                                               & \textbf{29.00}             & 58.54            \\ \hline
\multicolumn{3}{|c|}{includ. seq. mean-teacher step}                                                                  & 28.16             & 61.23            \\ \hline
\multicolumn{3}{|c|}{\begin{tabular}[c]{@{}c@{}c@{}}includ. seq. mean-teacher step \\ and supervised learning  \\ with seq. labels \end{tabular}} & 28.97             & \textbf{65.99}            \\ \hline
\end{tabular}%
}
\end{table}

\subsubsection{Sequential labeling for semi-supervised learning}
\label{sec:method2}
Table \ref{tab:Method2} shows the F-scores on the mean-teacher-based approach for semi-supervised learning with the DCASE 2019 Task 4 Training set for training. 
Including the sequential mean-teacher step in the conventional approach improved the segment-based F-score with a slightly degraded event-based F-score because the early step with sequential predictions mitigated inaccurate strong predictions and focused on occurrences of events rather than their timestamps. 
Furthermore, adding supervised learning with sequential labels to the previous approach improved both the event- and segment-based F-scores with significant improvement on the segment-based F-score, which was consistent with the results in Table~\ref{tab:M1_result_woMT}. In particular, both the sequential labels and predictions were helpful for improving the segment-based F-score by focusing on occurrences of events consistently. 

\section{Conclusion}
In this paper, we have proposed to use sequential information for mitigating label noise and inaccurate prediction in an early step of semi-supervised learning. Through the experiments on the recent public datasets, we demonstrated that using sequential information could improve AED performance.



\bibliographystyle{IEEEtran}

\bibliography{mybib}


\end{document}